# Dimensional-Crossover-Driven Mott Insulators in SrVO$_3$ Ultrathin Films

Man Gu[1, a)], Stuart A. Wolf[1, 2] and Jiwei Lu[2, b)]

[1]Department of Physics, University of Virginia, 382 McCormick Rd., Charlottesville, VA 22904

[2]Department of Materials Science and Engineering, University of Virginia, 395 McCormick Rd., Charlottesville, VA 22904

Author to whom correspondence should be addressed:

[a)] E-mail: mg8yd@virginia.edu

[b)] E-mail: jl5tk@virginia.edu

**Abstract:**

High-quality epitaxial SrVO$_3$ (SVO) thin films of various thicknesses were grown on (001)-oriented LSAT substrates by pulsed electron-beam deposition technique. Thick SVO films (~25 nm) exhibited metallic behavior with the electrical resistivity following the $T^2$ law corresponding to a Fermi liquid system. We observed a temperature driven metal-insulator transition (MIT) in SVO ultrathin films with thicknesses below 6.5 nm, the transition temperature $T_{MIT}$ was found to be at 50 K for the 6.5 nm film, 120 K for the 5.7 nm film and 205 K for the 3 nm film. The emergence of the observed MIT can be attributed to the dimensional crossover from a three-dimensional metal to a two-dimensional Mott insulator, as the resulting reduction in the effective bandwidth $W$ opens a band gap at the Fermi level. The magneto-transport study of the SVO ultrathin films also confirmed the observed MIT is due to the electron-electron interactions other than localization.

The metal-insulator transition (MIT) in 3$d$ transition metal oxides (TMO) has been a topic of long-standing interest in condensed matter physics.[1] Materials undergo a purely electronic MIT without structural changes are of particular interest, the basic mechanisms include Mott transition[2,3] and Anderson localization[4], which are based on electron-electron interactions and disorder-induced electron localization, respectively. Although the two routes can both separately induce a MIT, disorder is inevitably present in realistic strongly correlated systems, the subtle interplay of electron correlation and localization effects makes it challenging to determine the dominant driving force behind the transition.[5,6] Searching for new systems exhibit MITs and unraveling the transition mechanisms will open new avenues to predict, understand and control MIT, which could potentially impact the emerging field of oxide electronics.[7]

SrVO$_3$ (SVO) with a 3$d^1$ electronic configuration for vanadium is a typical strongly correlated system for studying MIT. SVO crystallizes in a cubic perovskite structure with a lattice constant of 3.843 Å.[8] Bulk SVO has been reported to be on the metallic side of a MIT with electrical resistivity ranging from $10^{-5}$ to $10^{-3}$ Ωcm at room temperature.[8,9,10,11] According to the Hubbard model, the control parameters for a Mott transition can be classified into bandwidth control and band-filling control.[1] In the bandwidth-controlled MIT (BC-MIT), the system is determined by the competition between the on-site Coulomb repulsion $U$ (localizes the electron), and the one-electron bandwidth $W$ (the tendency of electrons to delocalize), when $U/W$ is beyond a critical value, the system becomes a Mott insulator. In the filling controlled MIT (FC-MIT), the chemical doping of a Mott insulator with a half-filled band (filling $n = 1$) results in non-integer $n$ and a transition to a metal. Most previous studies of MIT in bulk SVO were achieved by chemical substitution. It has been demonstrated that a FC-MIT can be induced in the La$_{1-x}$Sr$_x$VO$_3$ system via aliovalent A-site substitution.[12,13,14] Also, the study of the metallic Ca$_{1-}$ $_{x}$Sr$_x$VO$_3$ system near the MIT has shown that the bandwidth can be tuned by isovalent A-site substitution with different ionic radius.[15,16] However, chemical substitution would unavoidably introduce significant disorder into the system, making it difficult to isolate the purely electronic contributions from the chemical disorder effects.

In recent years, improvements in thin-film correlated oxides synthesis techniques have provided new methods to study the MIT in ways not possible with bulk materials. Using these techniques, SVO thin films have been investigated, and the transport studies all showed good metallic behavior comparable with bulk SVO.[17,18,19,20] Most importantly, a dimensional-crossover-driven MIT in SVO ultrathin films has been reported recently using *in situ* photoemission spectroscopy (PES), it has shown that the observed MIT was induced by a reduction in the bandwidth due to the dimensional crossover when reducing the film thickness to 2-3 monolayers (ML).[21] The angle-resolved photoemission spectroscopy (ARPES) measurements of SVO quantum well states with film thicknesses ranging from 10 ML to 5 ML also confirmed the dimensional-crossover-driven MIT.[22,23] In this study, we report the electrical transport properties of SVO thin films as a function of film thickness, the temperature dependence of electrical resistivity revealed a MIT induced in SVO ultrathin films with thickness less than 6.5 nm (~17 ML).

High-quality epitaxial SVO thin films with thicknesses ranging from 3 nm to 25 nm were deposited on (001)-oriented (LaAlO$_3$)$_{0.3}$(Sr$_2$AlTaO$_6$)$_{0.7}$ (LSAT) (a = 3.868 Å) substrates using a pulsed electron-beam deposition (PED) technique. The base pressure of the PED system (Neocera Inc.) was ~5×10$^{-8}$ Torr. The SVO thin films were grown at a substrate temperature of 800 °C in a 10 mTorr Ar atmosphere. A 12 kV potential in the electron gun was applied to ablate a ceramic target of Sr$_2$V$_2$O$_7$ at a rate of 5 Hz. The film thicknesses were determined by using x-ray reflectivity (XRR) and the deposition rate was found to be ~2 Å/min. The crystalline structure of the films was examined by x-ray diffraction (XRD) (Smartlab, Rigaku Inc.) using Cu Kα radiation. The surface morphology was analyzed by atomic force microscopy (AFM) (Cypher, Asylum Research Inc.). The transport measurements were carried out in the temperature range of 2-300 K using a Physical Property Measurement System (PPMS) (Quantum Design Inc.), the electrical transport properties were measured using the van der Pauw method with cold-welded indium contacts on the 5 × 5 mm² square shaped samples.

As shown in Fig. 1a, out-of-plane XRD scans confirmed that the thick SVO film was single phase, and the Kiessig fringes around (002) symmetric reflection indicated smooth and coherent SVO films with a high-quality SVO/LSAT interface, the measured out-of-plane lattice parameter was ~3.846 Å. The reciprocal space mapping on the ($\bar{1}$03) asymmetric reflection in Fig. 1b also confirmed coherent growth of the SVO thin films with thicknesses up to 25 nm (~65 ML). The growth of SVO films on LSAT substrates was carried out using a layer-by-layer growth mode,[24] giving it a small lattice mismatch of -0.65%. In Fig. 1c, a typical AFM topography image of an SVO film showed clear step-edges and an atomically flat film surface with RMS roughness of ~0.2 nm. Also, the inset AFM line scan indicated step-terrace structures with each step height close to one unit cell lattice spacing of ~4 Å.

The temperature dependence of electrical resistivity was measured in the temperature range of 2-300 K. As shown in Fig. 2a, the 25 nm SVO film exhibited metallic behavior, the resistivity was about $1.17 \times 10^{-4}$ Ωcm at room temperature and $7.37 \times 10^{-5}$ Ωcm at 2 K, and the resistivity ratio $\rho(300\ K)/\rho(2\ K)$ was ~1.6. The electrical resistivity as a function of temperature can be expressed by a $\rho = \rho_0 + AT^2$ relation corresponding to a Fermi liquid system, where the residual resistivity $\rho_0$ is a temperature independent value attributed to the electron-impurity scattering caused by defects, and the coefficient $A$ quantifies the electron-electron interactions in the strongly correlated system. The fitting yielded $\rho_0 = 7.35 \times 10^{-5}$ $\Omega$cm and $A = 4.95 \times 10^{-10}$ $\Omega$cm/K$^2$, which is consistent with the reported values of bulk SVO.[8] Hall effects investigated over the temperature range of 2-300 K are shown in Fig. 2b. The carriers were found to be electrons with a concentration of ~$3 \times 10^{22}$ cm$^{-3}$, which is on the same order of magnitude as the predicted value assuming a simple one-band model. As the temperature was lowered from room temperature, the electron mobility slightly increased from 1.9 cm$^2$/Vs at 300 K to 3.1 cm$^2$/Vs at 2 K, while the carrier concentration stayed relatively constant.

Fig. 3a shows the electrical resistivity as a function of temperature for SVO films with thicknesses of 3 nm to 25 nm. Reducing film thickness caused an increase in film resistivity in the entire temperature range. As discussed above, clear metallic behavior was observed in the 25 nm and 15.3 nm SVO films. For the SVO ultrathin films with thicknesses below 6.5 nm, the films exhibited a temperature driven MIT characterized by a resistivity upturn at a transition temperature $T_{MIT}$ (arrows in Fig. 3b-3d). The transport data of the SVO films in this study is listed in TABLE I. Above $T_{MIT}$, the resistivity as a function of temperature for the SVO ultrathin films below 6.5 nm (~17 ML) showed metallic behavior, the resistivity decreased with the decrease of temperature, which indicated that the film was still continuous when the thickness was reduced to 3 nm (~8 ML). In this metallic regime, the temperature dependence of resistivity still followed the $T^2$ law, an increase in the estimated values of $\rho_0$ and A was observed with decreasing film thickness. The resistivity reached a minimum at $T_{MIT}$. As the temperature was lowered further from $T_{MIT}$, the resistivity gradually rose up as the films became insulators. By reducing film thickness, the $T_{MIT}$ shifted to the higher temperature, it was found to be at 50 K for the 6.5 nm film, 120 K for the 5.7 nm film and 205 K for the 3 nm film.

The emergence of the MIT induced in SVO ultrathin films with thicknesses below 6.5 nm could be attributed to a dimensional crossover from a three-dimensional metal to a two-dimensional Mott insulator. In a BC-MIT, the system is determined by the $U/W$ ratio.[1] The bandwidth $W$ is proportional to the number of available orbital states and thus highly sensitive to the film thickness. The atomic coordination at the interface and surface decreases with decreasing film thickness, resulting in a reduction of the effective bandwidth $W$. According to the Hubbard model, a system with small $U/W$ (wide bandwidth $W$) corresponds to a metallic phase, in which the fully occupied lower Hubbard band (LHB) and the unoccupied upper Hubbard band (UHB) overlaps. As the bandwidth $W$ decreases, the splitting between LHB and UHB gradually increases, and opens what can be described as a pseudogap at $E_F$, the pseudogap reflects an intrinsic spatial inhomogeneity in which some parts of the films have opened up a gap while others have not. This pseudogap eventually evolves into a complete band gap corresponding to a Mott insulator. The observed temperature driven MIT induced in SVO ultrathin films shows an transition region as the SVO films change from metallic states to insulating states from the dimensional crossover. The transition region as mentioned before, may result from some small thickness inhomogeneity that results in both the metallic and insulating phases coexisting at the same temperature, the slightly thinner regions transforming before the thicker regions. This can be thought of as percolative behavior. As the overall thickness of the films was reduced this transition moved to a higher $T_{MIT}$.

To understand the driving force behind the observed MIT induced in the SVO ultrathin films, the electron localization effects also need to be considered. In the insulating regime of the SVO ultrathin films with thickness less than 6.5 nm, the temperature dependence of resistivity below 30 K can be scaled with Mott's variable range hopping (VRH) model (insets of Fig. 3b-3d), which might be associated with the disorder-induced localized states at the Fermi level, or the percolation hopping between the granules as the films became very thin.[25,26] However, the magneto-transport properties of the SVO ultrathin films are very different from the LaNiO$_3$ ultrathin films with localized electron states.[27] As shown in Fig. 4a, a small positive magnetoresistance (MR, defined as [$R(B)–R(0)$]/$R(0)$) in a perpendicular magnetic field was observed in the insulating regime of the SVO ultrathin film at the temperatures below $T_{MIT}$. The MR increased with decreasing temperature as the film became more insulating, and the MR was found to be proportional to the $B^2$ which can be attributed to the Lorentz contribution.[28] The observed positive MR is unexpected from electron localization, since the magnetic field suppresses the coherent interference for localization and results in a negative MR.[29,30] Also, the Hall measurements of the same SVO ultrathin film are shown in Fig 4b., the carriers were electrons as observed in the thick SVO film, as the temperature was lowered from room temperature, the electron mobility slightly increased while the concentration slightly decreased and stayed ~7 × 10$^{21}$ cm$^{-3}$ below 50 K. However, in a disordered interacting electron system, the Hall coefficient was found to increase logarithmically at a rate equal to twice that of the resistivity at low temperature,[30] the temperature dependence of the carrier concentration observed in the SVO ultrathin films reported here, did not at all exhibit that dependence. Thus, based on our magneto-transport study, it seems safe to conclude that the dominant driving force behind the MIT induced in SVO ultrathin films is electron-electron interactions other than localization.

The dimensional-crossover-driven MIT in SVO ultrathin films was first observed from *in situ* PES.[21] The PES spectra indicated a pseudogap formed at $E_F$ as reducing film thickness to 3-6 ML, and the pseudogap finally evolved into an energy gap at the film thickness of 2-3 ML. In their further study, the ARPES measurements of SVO quantum well states with film thickness ranging from 10 ML to 5 ML also confirmed the dimensional-crossover-driven MIT.[22,23] In our study, the temperature-driven MIT corresponding to the pseudogap at $E_F$ was observed with film thicknesses from 6.5 nm (~17 ML) to 3 nm (~8 ML), which is much thicker than the films showing pseudogaps in their work, this disagreement between the x-ray spectroscopy and transport measurements could be understood by considering the local metallic states in the global insulating region.[31,32] Similar dimensional-crossover-driven MIT has also been observed in $CaVO_3$ (CVO) thin films with thicknesses below 4 nm,[33] and also in the surface of single-crystal CVO and SVO with a thickness of a few nm, as a strong enhancement of $U/W$ was observed at the surface compared with the bulk.[34,35]

In conclusion, we have successfully synthesized high-quality epitaxial SVO thin films on LSAT substrates. The transport study showed that SVO thick films exhibited metallic behavior followed the $T^2$ law corresponding to the Fermi liquid model. The temperature driven MIT was induced in SVO ultrathin films with thicknesses less than 6.5 nm. The MIT may result from the reduction in effective bandwidth $W$ due to the dimensional crossover from a three-dimensional metal to a two-dimensional Mott insulator. The magneto-transport study of the SVO ultrathin films showing MIT also confirmed that the dominant driving force behind the observed MIT is electron-electron interactions other than localization.

**Acknowledgements:**

We gratefully acknowledge the financial support from the Army Research Office through MURI grant No. W911-NF-09-1-0398. The authors would like to thank Dr. Ryan Comes, Dr. Hongxue Liu and Yonghang Pei for their helpful discussions.

**References:**

[1] M. Imada, A. Fujimori, and Y. Tokura,  Rev Mod Phys **70**, 1039 (1998).

[2] N. F. Mott,  Proc. Phys. Soc. London, Sect. A **62**, 416 (1949).

[3] N. F. Mott, *Metal-Insulator Transitions*, 2nd ed. (Taylor and Francis, London, 1990).

[4] P. W. Anderson,  Phys. Rev. **109**, 1492 (1958).

[5] D. Belitz and T. R. Kirkpatrick,  Rev Mod Phys **66**, 261 (1994).

[6] E. Abrahams and G. Kotliar,  Science **274**, 1853 (1996).

[7] Z. Yang, C. Y. Ko, and S. Ramanathan,  Annu Rev Mater Res **41**, 337 (2011).

[8] P. Dougier, J. C. C. Fan, and J. B. Goodenough,  J Solid State Chem **14**, 247 (1975).

[9] B. L. Chamberland and P. S. Danielson,  J Solid State Chem **3**, 243 (1971).

[10] M. Onoda, H. Ohta, and H. Nagasawa,  Solid State Commun **79**, 281 (1991).

[11] V. Giannakopoulou, P. Odier, J. M. Bassat, and J. P. Loup,  Solid State Commun **93**, 579 (1995).

[12] F. Inaba, T. Arima, T. Ishikawa, T. Katsufuji, and Y. Tokura,  Phys Rev B **52**, R2221 (1995).

[13] S. Miyasaka, T. Okuda, and Y. Tokura,  Phys Rev Lett **85**, 5388 (2000).

[14] T. M. Dao, P. S. Mondal, Y. Takamura, E. Arenholz, and J. Lee,  Appl Phys Lett **99**, 112111 (2011).

[15] I. H. Inoue, I. Hase, Y. Aiura, A. Fujimori, Y. Haruyama, T. Maruyama, and Y. Nishihara,  Phys Rev Lett **74**, 2539 (1995).

[16] I. H. Inoue, O. Goto, H. Makino, N. E. Hussey, and M. Ishikawa,  Phys Rev B **58**, 4372 (1998).

[17] H. Ishiwara and K. Jyokyu,  Jpn J Appl Phys 2 **30**, L2059 (1991).

18  D. H. Kim, D. W. Kim, B. S. Kang, T. W. Noh, D. R. Lee, K. B. Lee, and S. J. Lee, Solid State Commun **114**, 473 (2000).

19  D. W. Kim, D. H. Kim, T. W. Noh, K. Char, J. H. Park, K. B. Lee, and H. D. Kim, J Appl Phys **88**, 7056 (2000).

20  J. A. Moyer, C. Eaton, and R. Engel-Herbert, Adv Mater **25**, 3578 (2013).

21  K. Yoshimatsu, T. Okabe, H. Kumigashira, S. Okamoto, S. Aizaki, A. Fujimori, and M. Oshima, Phys Rev Lett **104**, 147601 (2010).

22  K. Yoshimatsu, K. Horiba, H. Kumigashira, T. Yoshida, A. Fujimori, and M. Oshima, Science **333**, 319 (2011).

23  S. Okamoto, Phys Rev B **84**, 201305(R) (2011).

24  R. Comes, M. Gu, M. Khokhlov, H. X. Liu, J. W. Lu, and S. A. Wolf, J Appl Phys **113**, 023303 (2013).

25  O. Entin-Wohlman, Y. Gefen, and Y. Shapira, J Phys C Solid State **16**, 1161 (1983).

26  A. M. Goldman and S. A. Wolf, *Percolation, Localization, and Superconductivity*. (Plenum Press, New York and London, 1984).

27  J. Son, P. Moetakef, J. M. LeBeau, D. Ouellette, L. Balents, S. J. Allen, and S. Stemmer, Appl Phys Lett **96**, 062114 (2010).

28  J.M. Ziman, *Electrons and Phonons: The Theory of Transport Phenomena in Solids*. (Oxford University Press, USA, 2001).

29  G. Bergmann, Phys Rep **107**, 1 (1984).

30  B. L. Altshuler, D. Khmel'nitzkii, A. I. Larkin, and P. A. Lee, Phys Rev B **22**, 5142 (1980).

[31] E. Sakai, M. Tamamitsu, K. Yoshimatsu, S. Okamoto, K. Horiba, M. Oshima, and H. Kumigashira, Phys Rev B **87** (2013).

[32] Z. Sun, J. F. Douglas, A. V. Fedorov, Y. D. Chuang, H. Zheng, J. F. Mitchell, and D. S. Dessau, Nat Phys **3**, 248 (2007).

[33] M. Gu, J. Laverock, B. Chen, K. E. Smith, S. A. Wolf, and J. W. Lu, J Appl Phys **113**, 133704 (2013).

[34] K. Maiti, D. D. Sarma, M. J. Rozenberg, I. H. Inoue, H. Makino, O. Goto, M. Pedio, and R. Cimino, Europhys Lett **55**, 246 (2001).

[35] J. Laverock, B. Chen, K. E. Smith, R. P. Singh, G. Balakrishnan, M. Gu, J. W. Lu, S. A. Wolf, R. M. Qiao, W. Yang, and J. Adell, Phys Rev Lett, Accepted (2013).

TABLE I. Transport properties of SVO films with various thicknesses deposited on the LSAT substrates.

| Film Thickness | $\rho(300\ K)$ | $\rho(2\ K)$ | $T_{MIT}$ | $\rho(T_{MIT})$ | $\rho_0$ | $A$ |
|---|---|---|---|---|---|---|
| | ($\Omega$cm) | ($\Omega$cm) | (K) | ($\Omega$cm) | ($\Omega$cm) | ($\Omega$cm/K$^2$) |
| 25 nm (65 ML) | $1.17 \times 10^{-4}$ | $7.37 \times 10^{-5}$ | - | - | $7.35 \times 10^{-5}$ | $4.95 \times 10^{-10}$ |
| 15.3 nm (40 ML) | $2.30 \times 10^{-4}$ | $1.52 \times 10^{-4}$ | - | - | $1.51 \times 10^{-4}$ | $8.79 \times 10^{-10}$ |
| 6.5 nm (17 ML) | $3.41 \times 10^{-4}$ | $2.65 \times 10^{-4}$ | 50 | $2.59 \times 10^{-4}$ | $2.50 \times 10^{-4}$ | $9.93 \times 10^{-10}$ |
| 5.7 nm (15 ML) | $5.84 \times 10^{-4}$ | $5.59 \times 10^{-4}$ | 120 | $4.97 \times 10^{-4}$ | $4.62 \times 10^{-4}$ | $1.33 \times 10^{-9}$ |
| 3 nm (8 ML) | $1.08 \times 10^{-3}$ | $2.91 \times 10^{-3}$ | 205 | $1.02 \times 10^{-3}$ | $9.40 \times 10^{-4}$ | $1.49 \times 10^{-9}$ |

**Figure Captions:**

Fig. 1

(a) XRD scans around (002) reflection of SVO film with thickness of 25 nm showing Kiessig fringes. (Inset) Wide range scan showing that no secondary phases are present. (b) The reciprocal space mapping on ($\bar{1}03$) reflection of the same 25 nm SVO film. (c) AFM topography image (scan area: 3 × 3 µm$^2$) and (inset) line scan of SVO film with thickness of 6.5 nm.

Fig. 2

(a) Electrical resistivity vs. temperature for SVO film with thickness of 25 nm. A $\rho = \rho_0 + AT^2$ fit is also shown. (d) Temperature dependence of carrier density and mobility from Hall measurements of the same 25 nm SVO film.

Fig. 3

(a) Electrical resistivity vs. temperature for SVO films with thickness ranging from 3 nm to 25 nm. Temperature dependence of electrical resistivity in a different scale for SVO films with thicknesses of (b) 6.5 nm, (c) 5.7 nm and (d) 3 nm. Arrows indicate the local minima of resistivity. (Inset) Logarithm of conductivity as a function of $T^{-1/4}$ for the same (b) 6.5 nm, (c) 5.7 nm and (d) 3 nm SVO films. A linear fit (black line) to the data below 30 K is also shown.

Fig. 4

(a) Normalized out-of-plane MR of the 5.7 nm SVO ultrathin film between 5 and 50 K with the field up to 7 T. The MR is proportional to $B^2$, the parabolic fits at each temperature are also shown in solid lines. (b) Temperature dependence of carrier density and mobility from Hall measurements of the same 5.7 nm SVO ultrathin film.

Fig. 1

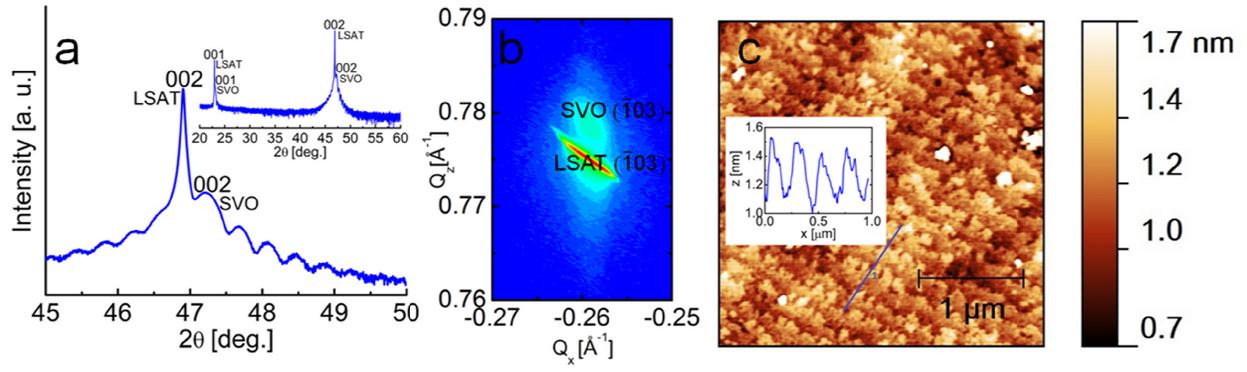

Fig.2

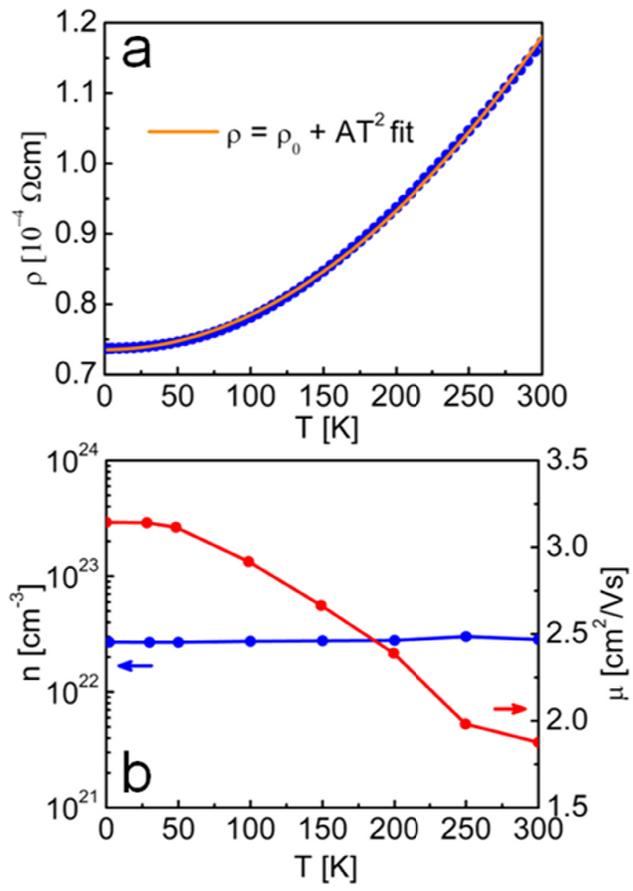

Fig. 3

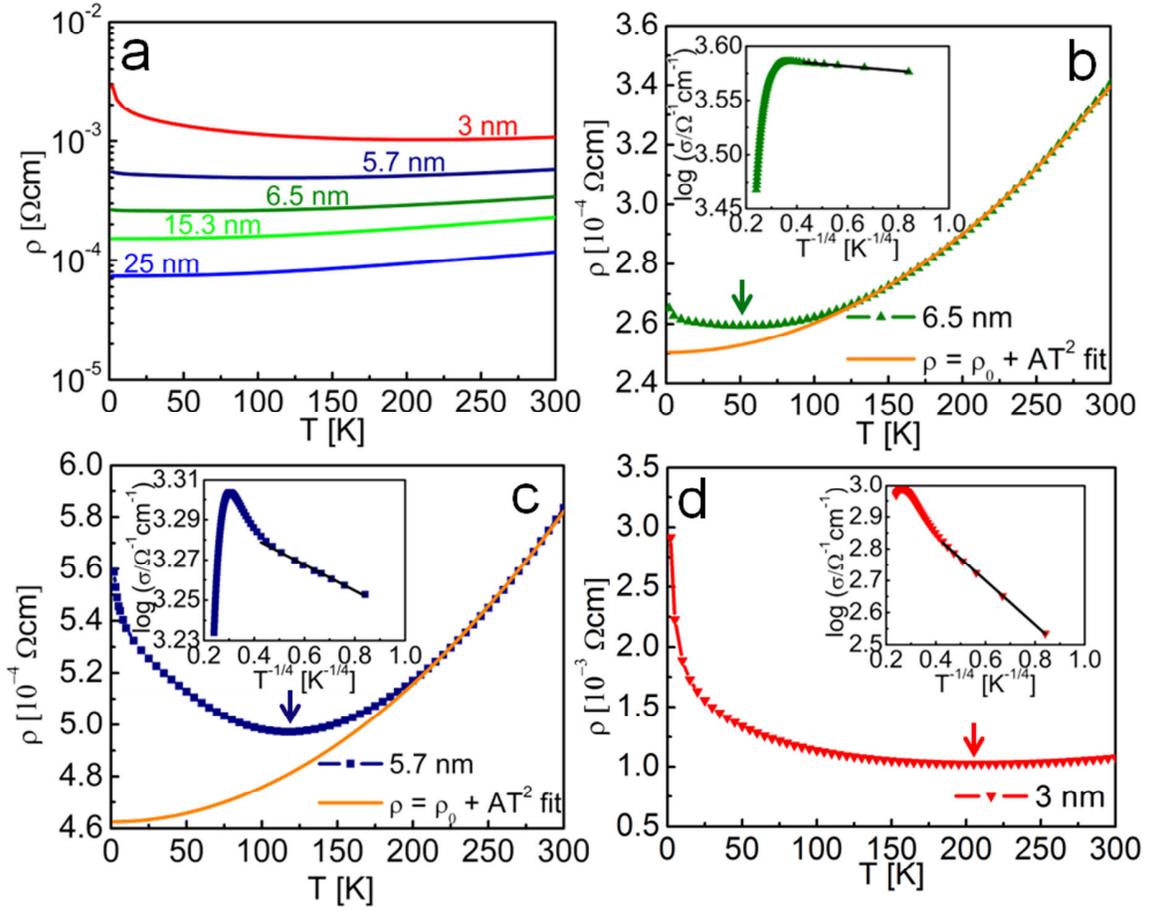

Fig. 4

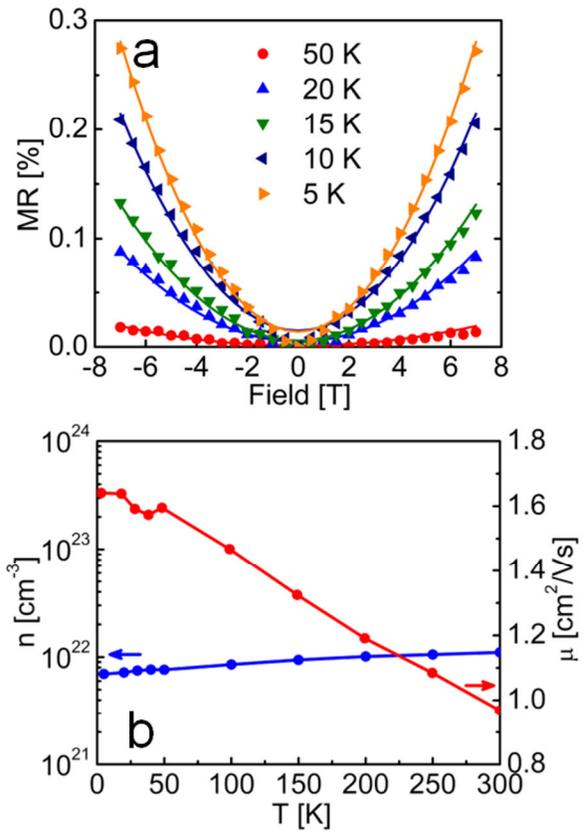